# Investigation the Critical Levels in Development of the Complex Systems with Shifted Arguments for their Optimal Control

KAZACHKOV[1,2] Ivan Vasilievich

[1]Royal Institute of Technology, Stockholm, Sweden; & [2]Nizhyn Gogol State University, Nizhyn, Ukraine

Investigation of the critical levels and catastrophes in the complex systems of different nature is useful and perspective. Mathematical modeling and analysis is presented for revealing and investigation of the phenomena and critical levels in a development of complex systems for various natures associated with diverse complicated factors, in particular with shifted arguments of the system. Intensive research in this direction and developed techniques may optimize management of the complex systems in financial-economic, natural and other fields. Construction of adequate mathematical models for development of complex systems, critical modes and their effective control are important tasks for a wide range of contemporary issues as shown in paper on examples. Critical levels in development of economic, banking, technical, political and other systems are necessary to determine and anticipate, to manage their system requirements and provide stable development, without being hit in a critical situations, leading to growing oscillations of the system settings.

## Introduction to the problem of systems' development and their harmonization

Mathematical modeling and computer simulation of the development of complex systems having various nature (from the living to the technical ones, financial, economic, etc., different by scales) associated with various complex factors, in particular with the presence of shifted arguments in the systems, have revealed critical processes and regimes in the contemporary area of research. Intensive research in this direction started in the last few decades and developed techniques to optimize the control of those systems, for example financial-economic, natural and others (Allen, 1976; Zhirmunsky & Kuzmin, 1990; Kazachkov, et al., 2004; Lee & Park, 2010; Gharakhanlou & Kazachkov, 2012; Kazachkov & Konoval, 2013).

The systems' development and their harmonization were in focus of many philosophers and scientists from different fields of science since ancient times, e.g. the Pythagoras adherents believed that planet motion is governed by the same numbers as the harmony of spheres (Fulier, 1901;) rhythmic unity of processes at diverse organizational levels were considered by ancient China and India (Sima Xjan, 1972; Mahabharata, 1978). Regularities of the critical levels, the Napierian number *e* as a module of the geometric progression, issues related to general problems of systems' development in different branches of science and practical activities were considered in a number of books and papers, e.g. the ones by Zhirmunsky & Kuzmin (1990), where a historical aspect was laid out together with the results of their own researches. The effective research method for a number of different developing systems has been elaborated and many examples of diverse origins, nature and scales have been analyzed in detail and amazing features revealed from just simple analysis, without solution of the equations.

Ivan V. KAZACHKOV[1,2]
[1]Dept of Energy Technology, Royal Institute of Technology, Stockholm, 10044, Sweden, Ivan.Kazachkov@energy.kth.se
[2]Dept of Applied Mathematics and Informatics, Nizhyn Gogol State University, Ukraine, http://www.ndu.edu.ua, kazachkov@ukr.net

Development and investigation of the physical and mathematical models for complex systems with shifted arguments and critical modes, their effective control are critical tasks for a wide range of modern problems, which are of paramount interest. Critical levels in a development of economic, banking, industry, technical, political and other systems, which may cause instability of the system and destroy a stable development of the system due to growing oscillations in a system, must be revealed and accounted in the strategic and tactical planning. Therefore the problem stated is vital for study, mathematical modeling and simulation. The theory of nonlinear dynamical systems with shifted (deviated) arguments provides a powerful mathematical tool for the study of complex systems and the determination of critical levels in their development. Properties of many real objects essentially depend on the after-effects due to which their behavior in the next moment of time depends on the previous history of development, and not only on the current state of the object.

The simplest cases of such systems have been studied in the theory of functional differential equations with shifted arguments (delay and forecast terms by time): Elsholz & Norkin (1971), Allen (1976), Baker, et al. (1995). Then during the recent times: Yan (2001), Rajeev, et al. (2002), Kazachkov, et al. (2004); in the $2009^{th}$ –Wu.; Liang, et al.; Björklund & Ljung; Jiang & Yang; Zhang, et al.; Sun & Song; in the next, $2010^{th}$ – Lee & Park; Wen, et al.; Pue-on & Meleshko; recently - Begun, et al. (2012); Kazachkov (2016).

It should be noted that real objects are more complicated and the mathematical models describing them, even with simplification described by systems of differential equations, contain the arguments depending on many deviating arguments, which, moreover, can themselves be time-dependent and to be linked: Zhirmunsky & Kuzmin (1990); Kazachkov, et al. (2004); Gharakhanlou & Kazachkov (2012); Gharakhanlou, et al. (2013, 2014). Over several decades the fundamental results in the theory of dynamical systems with delayed and forecasting arguments formed the theory of differential equations with shifted arguments applied during the last 30 years to modeling of complex systems from a wide range of science, technology, wildlife, economics and banking, and the like.

Development of numerical algorithms and their application to problem's solution devoted a lot of effort, e.g. Allen (1976), Elsholz & Norkin (1971), Baker, et al. (1995), Yan (2001) and Wu (2009), Jiang & Yang (2009), Zhang, et al. (2009), Lee & Park (2010), Wen, et al. (2010), Pue-on & Meleshko (2010), Azad, et al. (2002). But almost no attention was given to the equations with shifted arguments. Only Baker, et al. (1995) and Yan (2001) provided a classification task with forecasting arguments.

In relation to nonlinear dynamical systems with delayed and forecasting arguments, it should be noted that they were considered by Kazachkov, et al. (2004); Gharakhanlou & Kazachkov (2012); Begun, et al. (2012); Gharakhanlou, et al. (2013, 2014), Kazachkov (2016) for modeling of potentially hazardous objects of nuclear energy. Also in some papers the dynamics of populations crashing in biological systems and high voltage power lines were considered. Interestingly, in the theory of motion control with delay in time the application of necessary optimality conditions in the form of Pontryagin's maximum principle (Pontryagin, et al., 1961) leads to the conjugate system of the equations with forecasting arguments.

### Statement of the problem by mathematical modeling of complex systems

The effective numerical methods and methods of averaging the differential, integral, as well as the integro-differential operators allowing performing mathematical simulation in a wide range of complex processes and systems are applied for solution of differential equations with delayed and forecasting arguments. In particular, modeling the dynamics of behavior of potentially hazardous industries, based on statistical information about the objects. The aggregate models for the development of nuclear energy facilities were constructed and studied by Kazachkov I. V. with co-workers and PhD students. It allowed computational experiments to identify the interesting features in development of nuclear energy industry or an individual nuclear power plant, the optimal strategy, identification the critical and dangerous situations, etc. To some extent this can contribute to improving the management of the relevant objects and reduce their negative impact on the environment.

Since such complex objects in most cases do not allow constructing the precise deterministic mathematical models due to big or even huge number of influential parameters and often unknown links between them, then the aggregate model built on statistics about the object can be useful for studying the nature and behaviors of the objects. This paper focuses on the modeling and analysis of the behaviors of different systems based on the latest achievements in the theory, in particular, the method developed in the book by Zhirmunsky & Kuzmin (1990) who considered a lot of different applications as well.

The equations for the development systems with delayed and forecasting arguments are considered. It is shown, which levels must be the intensities of development and other parameters, so that the system is kept in a state of stable development and not subject to the modes of oscillations growing in time by amplitude that rapidly destroy it (instability, catastrophes, disintegration of the system). Our study focuses on peculiarities of the systems' behaviors in a methodological and mathematical aspects, the task of constructing and using such models of applied nature. Also the features performances of a number of outstanding tasks are discussed as well.

Development of the models for studying the evolution of complex systems (economic, political, social, banking, physical, combined by different nature and properties, etc.) in a rather general setting requires consideration of the delayed and forecasting temporal arguments, because development of the system is really accompanied by some delays compared to the planned indicators due to various reasons, as well as orientation for leading indicators what is known as the term "foreseen adaptation". Such phenomena have actually been observed in a number of different processes and systems as a wildlife and technical origin.

### Development of the mathematical models for complex systems

Probably first the equation for system development was already considered in 18$^{th}$ century by T. R. Malthus, English economist, statistician and demographer born at Guilford, Surrey (Wrigley, 1986; Keynes, 1963; Flew, 1970). Malthus was a pivotal figure in development of the empirical study of human populations who is very well known in the world for his *Essay on the Principle of Population*, the central theme of which was postulated inherent tendency for human numbers to `outstrip the means of subsistence', based on solution of the simplest development equation in the following form

$$\frac{dy}{dt} = ky, \qquad (1)$$

where *y,t* are the function describing evolution of some magnitude and time, respectively, *k* is the growth factor of the system, which generally can be function of time.

By Maltus and his followers human improvement therefore depended on stern limits on reproduction. The original essay, published anonymously in 1798, was directed against the optimistic speculations of the utopian writers believing in no limit to human progress. The solution of equation (1) for the simplest case with a constant growth rate *k* has shown that a law of population growth was seen as threatening to society and spawned the philosophical course of Malthusianism, which justified the war as a necessary mechanism for regulation of population growth. It was a big mistake based on a simplified equation of the system's growth. In fact, later on it was shown that the system can start development following the law (1), but further the coefficient $k$ depends on time, e.g., decreases according the hyperbolic law (so called allometric process). Thus, the function's increase with time decelerates. Since that time, there were many attempts to use this as a simple equation of development, and many others, more accurately given the characteristics of the systems under consideration.

The observation of a number of systems has shown the development processes better described in several other equations: over time, in some cases, coefficient $k$ is being gradually or abruptly falling down, and this leads to solutions, in which there is no exponential growth function. This behavior corresponds better to the realities of different systems of diverse nature: first there is an intensive development process, which is adjusted according to the results of the development and analysis of the development's needs. In the real systems some time delays and forecasting terms are met. The first ones are caused by delays in control mechanisms, and the second ones are available due to planning and orientation of the control system to the desired future performance rather than current real state.

### Nonlinear effects and nonlinearity due to time delay

The linearity of the systems' development may be broken in many cases (systems with more complex behavior), which in the example of equation (1) can be demonstrated as follows:

$$\frac{dy}{dt} = ky(A - y), \qquad (2)$$

where $A$ is the maximum possible value of $y$ under consideration. Limit values are defined by natural or artificial means in the specific system. For example, it may be a limit of a number of population for existing conditions with respect to given power supplies, the number of workers for the industry, if the industry is considered as a complex system and the limiting possible value is known, a limited amount of financial provision in development of the bank, and the like. In such cases, the fact that the closer functionality of a growth of a system to its limit is, the stronger growth rate of the system may naturally decrease. This is taken into account in the equation (2), and upon reaching the border, the further growth stops at the level $y=A$.

Another kind nonlinearity of the system (1) is available due to dependence of the coefficient $k$ against time and the function $y$ against deviating arguments:

$$\frac{dy}{dt} = ky(t - \tau), \qquad (3)$$

where $\tau$ is a value of time delay of the system. For solving the equation (3) the initial conditions on the time interval $\tau$ that precedes the starting point are needed, to specify in time or to examine the mechanism of delay (lag) only after a period of time $\tau$. This is a significant feature of the equations with time delay, which greatly complicates solution (the numerical solution of differential equations with automatic selection of time step must approximate point values with a shift in time, which may not be available ин automatic partitioning of the time interval). Equation (3) has much broader applications to modeling the development of complex systems because it accounts a possibility of time delays in the system's development relative to a current state of the system. The last one fits well to a number of real systems.

### Evolutionary and destroying regimes in the equations with the shifted arguments

By the positive growth factor $k$ in the equation (3), the solution of it is increasing function. If $k$ is negative –

attenuation system (decrease of the function *y*, that is, the drop development, extinction of populations, reducing the Bank's funding, etc. - depending on the nature of the complex system that is modeled). The equations with shifted arguments have more complex modes and features, in particular, critical stages of development and the possible instabilities that rapidly lead to a destruction of the system. Such regimes are particularly important and must be thoroughly investigated.

In complex real systems the equations of the type (3) can be many different and each of the plurality of interrelated system parameters may have its own delay, which significantly complicates the mathematical model of the system. The solution of (3) with the constant time delay $\tau$ and a growth factor $k$ can be sought in the form similar to a solution of the equation (1):

$$y = y_0 e^{zt} \qquad (4)$$

where $y_0$ is the initial value $y$ at $t = 0$, $z = u + iv$ are the Eigen values of the differential operator, *u, v* are, respectively, the real and imaginary part of the Eigen values, $i = \sqrt{-1}$ is the imaginary unit. By substitution of the requested solution in the form (4) into the equation (3) the equation for calculation of the Eigen values is got (after deletion by $e^{zt}$):

$$z = ke^{-z\tau}. \qquad (5)$$

Applying the Euler's formula for exponents of imaginary numbers to (5), for the real and imaginary parts of the equation (5) yields the following equation array:

$$u = ke^{-u\tau}\cos v\tau, \quad v = -ke^{-u\tau}\sin v\tau, \qquad (6)$$

where from follows: at *v=0* two processes are possible: monotonously exponentially increasing (*u>0*) and decreasing (*u<0*) in time.

When *v≠0*, the oscillating modes of system's development with exponentially decreasing (*u<0*) and exponentially increasing (*u>0*) amplitudes become available. In the first case, the oscillation process decays over time and is sustainable (can lead to degeneration of the system, the cessation of its operation), whereas in the second case, growing over time, the oscillation process will quickly destroy the system (growing in time instability cause catastrophe). For example, in a case of financial systems this means that the rocking of growth and decrease leads to a total collapse. So one needs to find the conditions, under which prevention of mode oscillations growing with time is possible (control of the system). Thus, it has to evolve, growing smoothly, without oscillations. From such general speculations one can come to conclusion that the financing of the project should be closely monitored for features of the development, which are modeled by the appropriate equations and scheduled possible delays in time.

If the time delay exceeds an upper limit for the system, there may be fluctuations in the parameters growing over time. The growing amplitude of oscillations may destroy the system. Therefore with the time delays in the development up to an extreme level nothing dangerous happens. But when the limits are exceeded, the whole system quickly collapses due to growing oscillations.

### Critical levels of the system and its control in the smooth pre-crisis regime

The study of the equations (6) by *v≠0* concerning an occurrence of values *u>0* leads to the conditions: $v\tau = (1.5 + 2n)\pi$, $k\tau = v\tau = (1.5 + 2n)\pi$, where *n= 0, 1, 2,...* Analysis shows that the first critical value, causes fluctuations of the system with growing amplitude is *kτ=3π/2*. Therefore, from (6) is got: $u\tau = 1.5\pi e^{-u\tau}$, which leads to the numerical solution $u\tau = 1.293$. Thus, a stable development of the system is possible only to certain limits, and a quick breakdown of the system due to growing fluctuations is coming. To prevent this, exponential growing must be managed continuing to achieving the time delay $\tau = 1.293/u$.

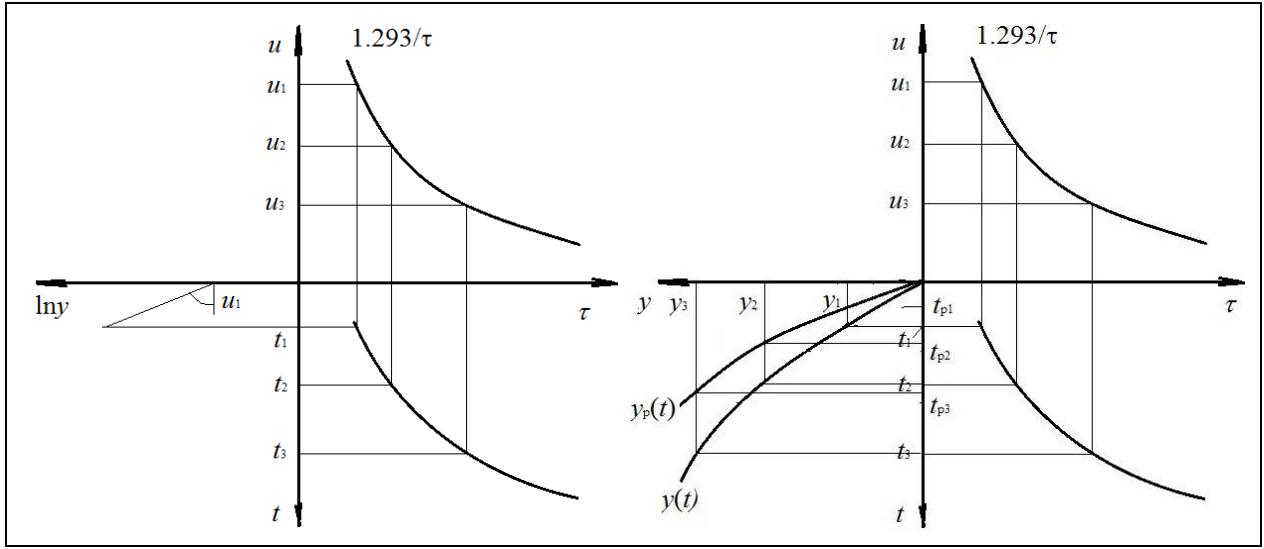

*Figure 1.* Critical levels of the growth rate $u$ versus time delay $\tau$ and corresponding values $y$ and time delay against time (left) and comparative characteristics of the real system's development $\tau$ and forecasting system without delays (right picture)

This means for a given $u$ the constant increase in time delay is permitted only up to a critical value $\tau = 1.293/u$ followed by rapid destruction of the system.

Based on the above mentioned, the strategy for sustainable development of the system with time delay requires managing the system, starting from any value of the growth rate of the system $u = u_1$, once it attains a critical value of delay $\tau = \tau_1 = 1.293/u_1$, the system's control must reduce the time delay or reduce the rate of the system's development. These features can be clearly traced in Figure 1 done according to Zhirmunsky & Kuzmin (1990). Here in figure to the right the critical levels of the growth rate of the system and associated delay in the first quadrant show the critical dependence of the rate of time delay for the system's development. And in the fourth quadrant the dependence of the lag (time delay) is shown, while in the third – lny versus time. Regimes of stable development of the system according to Figure 1 are below the curve $u = 1.293/\tau$ (the boundary region for sustainable development).

### Prediction of the regimes of stable development

The development can be optimally predicted in the following way. If the growth of the system starts with the given desired rate of $u_1$ at the beginning of such system's grow rate, it may be maintained only until the time $t = t_1$, when the increasing delay of the system at the time $\tau_1(t_1)$ becomes critical (line $u = u_1$ crosses a critical curve $= \frac{1.293}{\tau}$) at the point $\tau_1(t_1)$. Then further growth of the system with the specified rate is impossible, provided the further growth with the same time delay. It is necessary to reduce the rate of growth of the system, for example, up to some value $u = u_2$, then there is an additional system resource regarding the increase in the delay to the intersection of the critical curve at a lower level of growth.

If possible, for sustainable development it is necessary to reduce the growth of time delay for the system, which in many cases is poorly controlled or not controlled at all. Under the case of uncontrolled time delay one can improve the situation and control the system reducing the growth rate every moment when the critical values of time delay are got. In reality, the shift points of the control system according to the above described scenario can be realized in the following way. For example, the Bank finances the project, which for the equation of development has been identified in a specified form, whose solution was found.

Assume that the function $y(t)$ for sustainable development is known. If one needs to take into account some possible time delays in the system and manage it for optimal development and lack of critical modes, which destroy the system, it is necessary to determine a delay by comparing the characteristics of the system in real-time calculated, as shown in Figure 1 to the right. The forecasted system's parameters $y_p(t)$ obtained without taking into account the possible time delay in comparison with such parameters $y(t)$ obtained for a given delay are compared to identical values, resulting in delays expressed by the following expressions: $\tau_1 = t_1 - t_{p1}$, $\tau_2 = t_2 - t_{p2}$, $\tau_3 = t_3 - t_{p3}$, and so on. Thus, it is possible to determine the actual delay, if the results of the behavior of a real system are known. This allows determining the optimal modes of development of the system and preventing appearance of the critical (catastrophic) situations.

### Revealing the important features of system's development

In the monograph by Zhirmunsky & Kuzmin (1990), in addition to theoretical issues a large number of examples of applications were considered. Some amazing features of the development of bright diverse fields' systems with time delays were revealed and analyzed in detail. In particular, the question concerning development strategy in the region below the critical one (below the critical curve, shown in Figure 1) to what extent should reduce time delay or the growing rate to remain sustainable exponential development was stated. How to rebuild the system so that it continued to grow?

It is a very important issue for many natural and technical systems. So, practice has shown that most financial and economic systems do not receive such control and after a period of intensive development get into a state of slow development or stagnation, or total destruction. This is because the laws of development for the systems with time delays or/and forecasting, with the manifestation of significant nonlinearities are too complex for comprehending without solid mathematical modeling and optimization of development strategy, without taking into account the most essential features of the system behaviors, the switching mechanisms of the adjustment systems for the future desirable characteristics.

Any control of complex systems has not only to consider the time lag, but it is also based on the forecast, and therefore the mechanism of adjustment of the managed system, its sustainability must be proactive. Therefore, in many cases the development management of the systems requires simultaneous and consistent taking into account both the time delay and the switching mechanisms of adjustment (adaptation) of the system under the future features. To study characteristic mechanisms of timing and their impact on sustainable development of systems it can be considered, for example, the following mathematical model:

$$\frac{dy}{dt} = ky(t + \tau), \quad (7)$$

where $\tau$ is the time forecasting term of the system.

Equation (7) means that the rate of development of the system described by mathematical model, that focuses not on current performance as in equation (1) and not on previous performance as in equation (3), but on a future performance. To solve equation (7), the initial conditions on the time interval from zero to $\tau$ must be specified. For example, funded project and its execution at each moment of time has a rate that does not match the current level of funding, and one that meets the future, a higher level, which is the head. This is a case of development in advance. Modeling the development of complex systems with forecasting terms has interesting features that are worth exploring and are used for optimal control of systems' development (processes).

### Solution of the equation depending on forecasting terms by time

Considering solution of the equation (7) in the form (4) and substituting the solution sought $y = y_0 e^{zt}$ in (7), the equation to determine the Eigen values (after reduction by $e^{zt}$) yields: $z\tau = k\tau e^{z\tau}$. Researching possible solutions of this equation yields the highest value $k\tau$, by which solution exists, under conditions:

$$u\tau = 1, \quad k\tau = 1/e, \qquad (8)$$

where it is clear that as similarly to equation with time delay, when $v=0$ there are two cases: exponentially increasing ($u>0$) and exponentially decreasing ($u<0$) in time. When $v\neq0$, the oscillating modes of development in a system with exponentially decreasing ($u<0$) and exponentially increasing ($u>0$) amplitudes take place. In the first case, the process of oscillation is damped over time and is sustainable (can lead to the degeneration of the system, the cessation of its operation). But in the second case, growing over time, the oscillations quickly destroy the system. Only here the conditions (8), in contrast to the case of delay, define the lower boundary of the steady growth of the system. The rates of development of the system, taking into account forecasting in time that are below $u = 1/\tau$, will lead to increasing time-oscillations of the system, leading to its destruction.

### Strategies of correlated time delay and time forecasting in the system development

The above considered forces to conclude that for sustainable development of the system both processes of the time delay and time forecasting must be correlated in a system development. This can be treated as the global law of optimal control for systems. From (8) and similar expressions earlier obtained for critical levels of delay systems follows that the ratio of the growth rate for the delayed type and forecasting type systems' development is approximately equal to 1.293. By Zhirmunsky & Kuzmin (1990), it is possible to recommend the following method for determining the development of the system and its critical points, illustrating the process according to Figure 2, where in the first quadrant are two critical curves, respectively, for strategies with a delay (top) and forecasting (bottom).

In any case, a stable development is possible only between these two strategies and the actual behavior of the system can be only inside the region between them. Example of building the control strategy for system with prevention of critical regimes in steady development is considered below. Given the peculiarities of the behavior of systems development with shifted arguments, one can construct a control strategy in the following way. Let at the initial moment of time ($t = 0$) the intensity of development equal to $u_0$, delay is absent ($\tau=0$), so the system according to (4) develops following the law

$$y = y_0 e^{u_0 t} \qquad (9)$$

at time $t = t_1$, where the delay is equal to $\tau_1(t_1) = 1.293/u_0$, then it must switch the rate of development to the border region, which is leading to the arguments $u_1 = 1/\tau_1 = u_0/1.293$.

Next, the system continues stable growth law (9) with the new rate $u_1$ point-in-time $\tau_2(t_2) = 1.293/u_1$ when the delay reaches a critical level of development for a strategy with delay (see Figures 1, 2). Then again, one can go to the lower critical curve, which corresponds to the development strategy with forecasting, that is $u_2 = 1/\tau_2 = u_1/1.293$. And so on to provide the desired level of system development. is clear that over time of the system's development with growing time delay such control becomes ineffective so that it's best to stop and then continue the development, starting it again with zero delay.

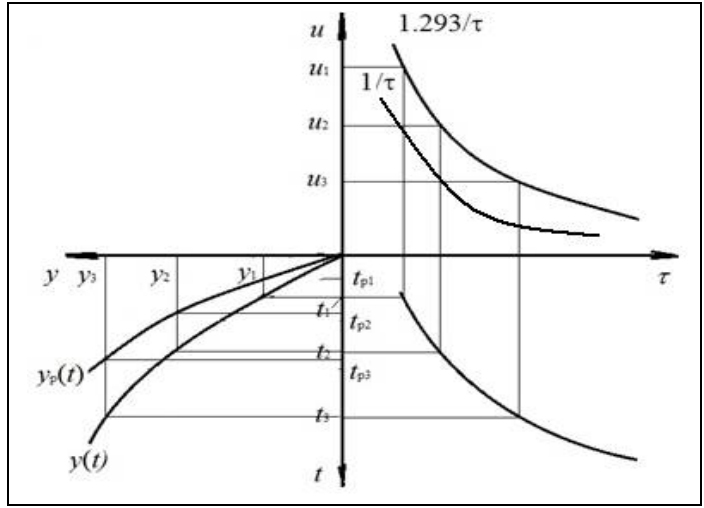

*Figure 2*. The scheme for stable development with correlated time delay and forecasting strategies

For the control of the time delay it is useful to have function $\tau(t)$, which can be obtained for example as a second order polynomial satisfying the conditions: $t = 0, \ \tau = 0; t = t_1, \tau = \tau_1; t = t_2, \tau = \tau_2$. Such function has a form

$$\tau = (a_1 + a_2 t)t, \quad a_1 = \frac{\tau_1 t_2^2 - \tau_2 t_1^2}{t_1 t_2(t_2 - t_1)}, \ a_2 = \frac{\tau_2 t_1 - \tau_1 t_2}{t_1 t_2(t_2 - t_1)}. \tag{10}$$

Hence, considering the above formula (10) implies the following expression for delay as a function of time:

$$\tau = \frac{1.293 t}{t_1 t_2 (t_2 - t_1)} \left[ \frac{t_1}{u_1}(t - t_1) + \frac{t_2}{u_0}(t_2 - t) \right]. \tag{11}$$

**Example of the stable development of income system**

The stable development of the system is according to (9) to the next critical point in time where the tempo switches to the next and so every critical moment. And the shift points of the development rate are calculated with (11). For example, $y_0 = 1$ bln USD, the initial growth rate of the financial plan adopted is $u_0 = 1.293/year$, so for the year according to the formula (9) needs to obtain $y = y_0 e^{1.293} \approx 3.63 y_0$, that is, to 3.63 bln income. But the time delay is $\tau_1 = 1$ year. So for next year one needs to change the growth rate to $u_1 = 1/year$. Over the next year the revenue should be about 2.71 bln. Having the delay equal to $\tau_2 = 1.293/year$, so for the time of 1.293 years income will be 3.63 bln. And then one needs to change the rate of development to the level of $u_2 = 1/1,293 \approx 0,77$. The latest rate gives an income of about 2.15 bln USD for the year, etc.

Many other critical levels can be investigated for diverse cases of the development systems the methodology developed in the book of Zhirmunsky & Kuzmin (1990). It is a topic for a few separate papers. The other, more complex real cases, which are not covered by their methods, are considered below.

## The nonlinear systems

Equation (1) is linear; equation (2) is nonlinear. The nonlinearity in the system by the mathematical model (2) is relatively simple. But the equation (3) with delayed arguments seems linear, but it contains the worst type of nonlinearity (2). If in equation (3) replace the function of delayed argument *y(t-τ)* according to theorem of Elsholtz (Elsholz & Norkin, 1971), which is satisfied for monotone functions, then the function *y(t-τ)* is represented in the Taylor series relative to the point *t* by *τ* with accuracy to the linear terms, since the linear approximation is the most precise in this case according to the theorem: $y(t - \tau) \approx y(t) - \tau dy/dt$, then equation (3) takes the following approximate form:

$$(1 + k\tau)\frac{dy}{dt} = ky(t). \tag{12}$$

Equation (12) differs relatively little from (1) and the delay affects only the time deformation. But when the function is non-monotonic, and such solutions can be oscillatory with increasing amplitude, the theorem of Elsholtz is broken, and instead the left side in (12) there will be a complete Taylor series to the derivative of the function *y* and the powers of delay *τ*. The last case is exactly critical, examples of which have just been considered. Significantly stronger effect of the time delay in case of the nonlinear equation (2) is:

$$\frac{dy}{dt} = ky(t - \tau)(A - y(t - \tau)), \tag{13}$$

which even for the case of monotonous growing, when Elsholtz theorem satisfies, leads to the strong nonlinear equation of the form

$$(1 + Ak\tau - 2k\tau y)\frac{dy}{dt} + \left(\tau \frac{dy}{dt}\right)^2 = ky(A - y), \tag{14}$$

containing the nonlinear terms of different type.

The nonlinearity of the systems and processes always cause unpredictable properties and characteristics of their behavior including the existence of various special and critical parameters and system modes. Possible points of bifurcation in a system correspond to the situations, where the system abruptly jumps from one mode to another (usually completely different from the previous one). There are available also strange attractors (sets of trajectories in the phase space of the system in which all other trajectories approach under any initial conditions), and the other features. The study of nonlinear properties of systems (13), (14) in many cases is crucial for understanding their behavior, which can be highly unpredictable. So, knowing the basic critical modes and the corresponding parameters and laws of management, one can try to optimize the system and to prevent the ingress of an object into a critical state.

## Complex systems with a number of parameters and shifted arguments

Complex systems with a large number of governing parameters and many possible delays and forecasting terms are difficult to investigate in the above-mentioned way. But the general patterns are similar on a qualitative level and must be accounted. For example, the aggregated mathematical model of a potentially hazardous object in nuclear energy (like the mathematical model can be developed also for other objects) has the following form (Kazachkov, et al., 2004):

$$\frac{dz_4}{dt} = [b_{40} + b_{43}z_3(t - \tau_{43}) + b_{44}z_4(t - \tau_{44}) + b_{45}z_5(t - \tau_{45}) + b_{46}z_6(t - \tau_{46})]z_4(t - \tau_{40}),$$

$$\frac{dz_1}{dt} = [b_{10} + b_{11}z_1(t-\tau_{11}) + b_{12}z_2(t-\tau_{12}) + b_{13}z_3(t-\tau_{13})]z_1(t-\tau_{10}),$$

$$\frac{dz_2}{dt} = [b_{20} + b_{21}z_1(t-\tau_{21}) + b_{22}z_2(t-\tau_{22}) + b_{23}z_3(t-\tau_{23})]z_2(t-\tau_{20}), \quad (15)$$

$$\frac{dz_3}{dt} = [b_{30} + b_{31}z_1(t-\tau_{31}) + b_{32}z_2(t-\tau_{32}) + b_{33}z_3(t-\tau_{33}) + b_{34}z_4(t-\tau_{34}) + b_{35}z_5(t-\tau_{35}) + b_{36}z_6(t-\tau_{36})]z_3(t-\tau_{30}),$$

$$\frac{dz_5}{dt} = [b_{50} + b_{53}z_3(t-\tau_{53}) + b_{54}z_4(t-\tau_{54}) + b_{55}z_5(t-\tau_{55}) + b_{56}z_6(t-\tau_{56})]z_5(t-\tau_{50}),$$

$$\frac{dz_6}{dt} = [b_{60} + b_{61}z_1(t-\tau_{61}) + b_{62}z_2(t-\tau_{62}) + b_{63}z_3(t-\tau_{63}))]z_6(t-\tau_{60}),$$

where $\tau_{ij}$ is the time delay for the corresponding parameters. Such indexes in a more general case also, in turn, depend on the time (and possibly each other). Here $z_i$- system parameters, $b_{ij}$- coefficients of the mathematical model, which are determined for each model based on the results of its functioning using the methods of identification. The system of differential equations (15) discussed above is much more difficult than equations of the type (13), (14) and allows the analysis of critical levels by described methods. It was investigated numerically. From the Figures 3-5 it is seen that modes can be critical and catastrophic, and detection of the parametric dependencies and characteristics is much more complicated than the above, where the concept of the main features is clearly understandable.

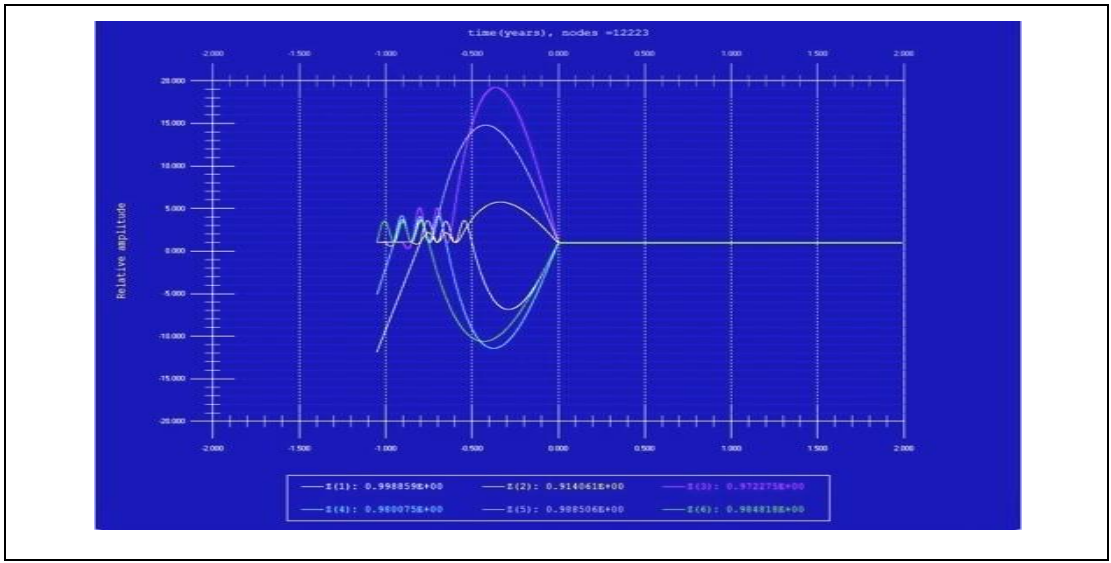

*Figure* 3. Oscillation processes in potentially hazardous object's development

**Examples of computer simulations for potentially hazardous objects**

The developed aggregated mathematical models of complex objects for a variety of applications, taking into account the effect of deviating arguments and nonlinear effects, allowed investigating some features of the objects by computer simulation. So, it is possible to set some negative impact on the environment and people, possible ways of weakening the negative actions and eliminate them, to determine the features of development dynamics of the object, change the safety culture on it and possibilities to improve it. Construction and research of aggregated dynamic models on a computer can be useful also for the study of tactical and strategic management of facilities at various levels (enterprise, industry or simply complex technical devices, etc., e.g. nuclear power plant Kazachkov, et al. (2012).

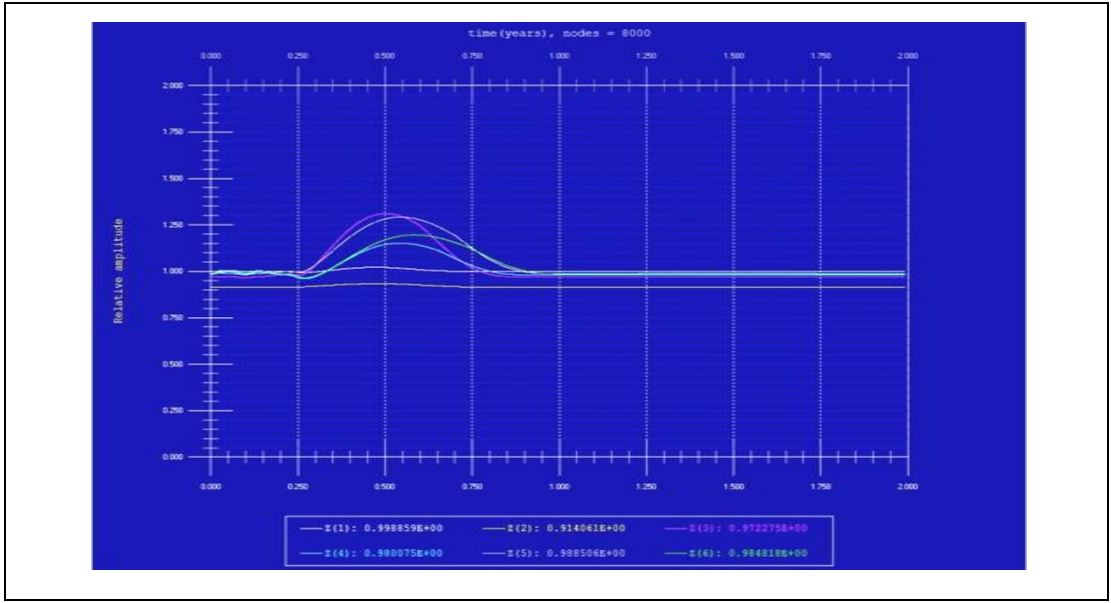

*Figure* 4. Oscillation processes in the development of aggregate nuclear power energy object

The results of computer situational simulations in a wide range of variable parameters given the shift of the arguments enable to identify and explore some of the most important features and phenomena as an optimizing plan, and dangerous for the development and operation of the facility. The identified critical conditions and catastrophic situations and scenarios for which the system can get to them, should be studied carefully with the purpose of their conditions in a real object, which is investigated by simulation.

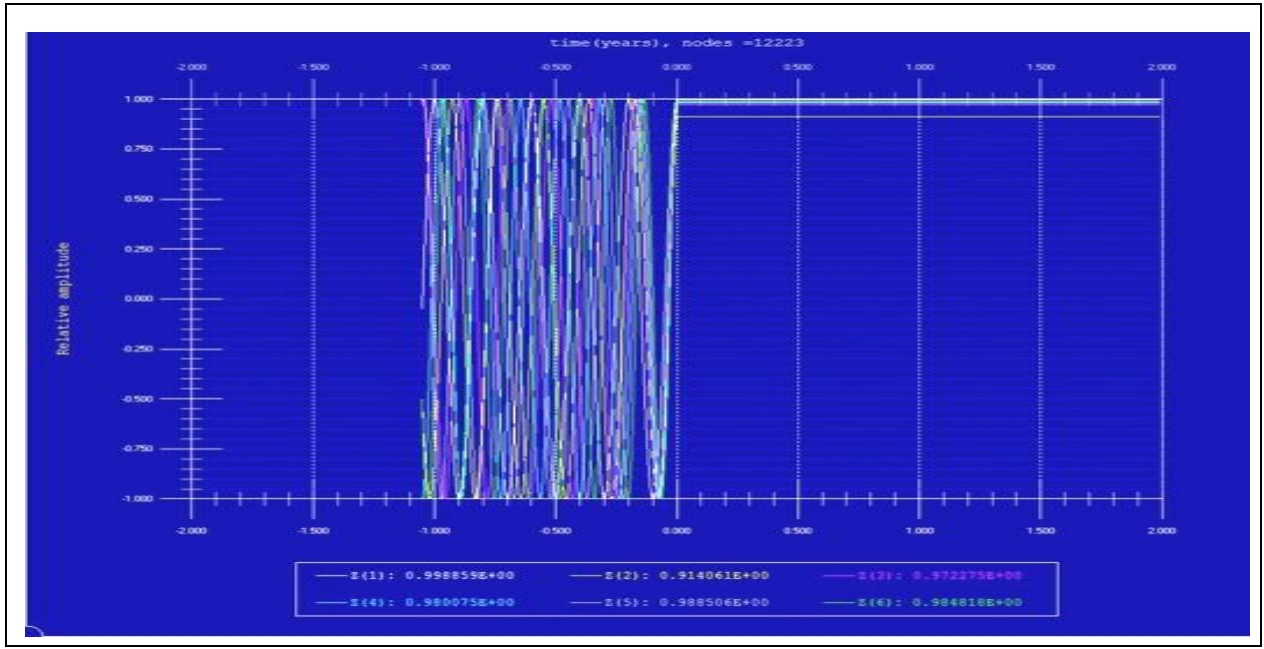

*Figure* 5. Functions $z_i$ (for *t>0*) and interpolations for interval till the initial moment (*t <0*)

## Conclusion

The approaches for development of the systems with shifted arguments and their implementation, modeling methodology and analysis of the obtained data showed how knowledge about the dynamics of development allows systematizing the data about the critical situations and mutual influence of different parameters, in which a variety of undesirable or even catastrophic events become possible. Example on development of financial system showed a variation of the management policies between the delays and forecasting based on piecewise continuous matching. The development starts from the selected tempo, which is maintained until achieving a critical level with increasing delay over time, and then switches to mode critical curve of the trajectory, which corresponds to strategy of development ahead. And so it's constantly switching between strategies with delay and forecast.